\documentclass[conference,10pt,a4paper]{IEEEtran}
\usepackage{amsthm}
\newtheorem{theorem}{Theorem}[]
\newtheorem{lemma}[theorem]{Lemma}

\newtheorem{example}{Example}

\usepackage{environ}

\NewEnviron{problem}[1]{%
	\begin{center}\fbox{\parbox{0.95\columnwidth}{%
				{\centering\scshape #1\par}%
				\parskip=1ex
				\everypar{\hangindent=1em}%
				\BODY
			}}\end{center}}

\usepackage{cite}

\usepackage{amsmath}
\usepackage{amssymb}

\usepackage[vlined, ruled, boxed, linesnumbered]{algorithm2e}
\SetKwInput{KwData}{Input}
\SetKwInput{KwResult}{Output}

\makeatletter
\newcommand{\longdash}[1][2em]{%
  \makebox[#1]{$\m@th\smash-\mkern-7mu\cleaders\hbox{$\mkern-2mu\smash-\mkern-2mu$}\hfill\mkern-7mu\smash-$}}
\makeatother
\newcommand{\omitskip}{\kern-\arraycolsep}

\usepackage{color}

\newcommand{\genMat}{\mathbf{G}}
\newcommand{\genVec}{\mathbf{g}}
\newcommand{\obsVec}{\mathbf{x}}
\newcommand{\obsMat}{\mathbf{X}}
\newcommand{\codeSet}{\mathcal{C}}
\newcommand{\code}{C}
\newcommand{\codeVec}{\mathbf{c}}

\begin{document}

\bstctlcite{IEEEexample:BSTcontrol}

\title{On the Computational Complexity of Blind Detection of Binary Linear Codes}

\author{\IEEEauthorblockN{Alexios Balatsoukas-Stimming}
\IEEEauthorblockA{Department of Electrical Engineering\\
	{\'E}cole polytechnique f{\'e}d{\'e}rale de Lausanne\\
CH-1015 Lausanne, Switzerland\\
Email: alexios.balatsoukas@epfl.ch}
\and
\IEEEauthorblockN{Aris Filos-Ratsikas}
\IEEEauthorblockA{Department of Computer Science\\
	{\'E}cole polytechnique f{\'e}d{\'e}rale de Lausanne\\
CH-1015 Lausanne, Switzerland\\
Email: aris.filosratsikas@epfl.ch}}


\maketitle

\begin{abstract}
In this work, we study the computational complexity of the \textsc{Minimum Distance Code Detection} problem. In this problem, we are given a set of noisy codeword observations and we wish to find a code in a set of linear codes $\codeSet$ of a given dimension $k$, for which the sum of distances between the observations and the code is minimized. We prove that, for the practically relevant case when the set $\codeSet$ only contains a fixed number of candidate linear codes, the detection problem is NP-hard and we identify a number of interesting open questions related to the code detection problem.
\end{abstract}


%
\IEEEpeerreviewmaketitle

\section{Introduction}\label{sec:intro}

Modern communications systems usually employ adaptive modulation and coding (AMC) mechanisms to cope with the highly varying channel conditions. In an AMC scenario, the devices at the two endpoints of each communication link agree on a combination of modulation and coding through a control channel. However, in recent communications standards, the control channel can itself use one of several modulation and coding combinations. It has thus become essential for wireless devices to be able to blindly detect and decode the information on the control channel in order to successfully join the wireless network. In practice, several parameters may need to be blindly detected (e.g., modulation, coding, interleaving), but in this work we focus on the problem of blind channel code detection, which can be loosely formulated as follows. Given a set of candidate codes $\codeSet$, a set of noisy codewords, and the knowledge that all of the noisy codewords are produced by the same code $\code \in \codeSet$, what is the most ``plausible'' candidate code $C \in \codeSet$ to have generated those words? 

The design of practical algorithms for the above version of blind detection of channel codes has drawn significant attention in the past years. For example, various heuristic methods have been proposed for the blind detection of Hamming and BCH codes~\cite{Yardi2014,Chabot2007}, convolutional codes~\cite{Cluzeau2009,Moosavi2011}, Turbo codes~\cite{Debessu2012,Tillich2014}, LDPC codes~\cite{Xia2014,Yu2016}, and polar codes~\cite{Condo2018, Condo2017, Giard2017, Giard2018}. In contrast, comparatively little is known about the fundamental computational complexity of the blind code detection problem.

\subsubsection*{Contributions} To the best of our knowledge, this is the first work that \emph{formally} studies the computational complexity of the blind code detection problem. To this end, in Section~\ref{sec:background} we first express the problem in a form that enables us to theoretically analyze its computational complexity. Then, in Section~\ref{sec:mdcd} we examine the practically relevant case where $\codeSet$ contains only a constant number of candidate linear codes (i.e., $|\codeSet| = \ell,~\ell > 0$) and we show that the \textsc{Minimum Distance Code Detection} problem in this case is NP-hard. In essence, our hardness result justifies the heuristic approach of a large body of existing work (c.f., \cite{Yardi2014, Chabot2007, Cluzeau2009, Moosavi2011, Debessu2012, Tillich2014, Xia2014, Yu2016, Condo2018, Condo2017, Giard2017, Giard2018} and references therein). In the related work of~\cite{Valembois2001}, the author formulated the problem when $\codeSet$ is the set of all linear codes of dimension $k$. While this choice of $\codeSet$ is appropriate for some scenarios (cf.~\cite[Sec. I]{Carrier2018}), the case where $|\codeSet| = \ell$ is much more natural and has a greater practical significance, since in most applications~(cf. \cite{Yardi2014, Chabot2007, Cluzeau2009, Moosavi2011, Debessu2012, Tillich2014, Xia2014, Yu2016, Condo2018, Condo2017, Giard2017, Giard2018}) the set of candidate codes is usually small and pre-defined by the employed communication standard. We discuss the relation between~\cite{Valembois2001} and our work in more detail in Section~\ref{sec:valembois}. Finally, in Section~\ref{sec:open} we identify and discuss a number of interesting related open problems.

\section{Blind Code Detection Background}\label{sec:background}
In this section, we first provide some brief background on binary linear codes and we then define the \textsc{Minimum Distance Code Detection} problem.

\subsection{Binary Linear Codes}
A binary linear code $\code$ of length $n$ is a set of $n$-bit vectors, called \emph{codewords}, with the property that for any $\codeVec_1, \codeVec_2 \in \code$, we also have $\codeVec_1 + \codeVec_2 \in \code$, where additions are performed using {modulo-$2$} arithmetic. The dimension $k$ of the code $\code$ is equal to the dimension of the subspace spanned by the codewords in $\code$. The number of codewords of a binary linear code of dimension $k$ is $2^k$. A binary linear code $\code$ can be efficiently represented using a $k \times n$ binary generator matrix $\genMat$ of rank $k$, so that each codeword can be generated as $\mathbf{u}\genMat$, for some ${\mathbf{u} \in \{0,1\}^k}$, and where all operations are carried out using {modulo-$2$} arithmetic. We use $\text{span}(\genMat)$ to denote the row span of $\genMat$, i.e., $\text{span}(\genMat) = \left\{\mathbf{u}\genMat: \mathbf{u} \in \{0,1\}^k\right\}$. Note that $\code = \text{span}(\genMat)$ and, due to this equivalence, we slightly abuse the terminology for simplicity and we refer to $\genMat$ both as a \emph{generator matrix} and as a \emph{code} depending on the context.

\subsection{Minimum Distance \& Maximum Likelihood Code Detection}
The blind detection problem can be formally stated as follows. Let $\obsVec_1, \ldots, \obsVec_N$, denote a set of $N$ binary row vectors of length $n$ that are observed at the output of a noisy channel and let the matrix $\obsMat$ be defined as:
\begin{align}
	\obsMat	& = \begin{bmatrix}
		\obsVec_{1}^T & \hdots & \obsVec_{N}^T
	\end{bmatrix}^T.
\end{align}
We will refer to $\obsVec_1, \ldots, \obsVec_N$ as the \emph{noisy codewords} and to the matrix $\obsMat$ as the \emph{observation matrix}. The code detection problem can generally be defined as follows. Given a set of codes $\codeSet$, an observation matrix $\obsMat$, and the knowledge that all of the noisy codewords are produced by the same code in $\codeSet$, find a code $C \in \codeSet$ that optimizes an appropriately defined metric. We briefly describe two distinct code detection problems that use different metrics below.

In \textsc{Minimum Distance Code Detection} (MDCD) the goal is to minimize the sum minimum distance between the noisy codewords in $\obsMat$ and the code $\code$. More specifically, let:
\begin{align}
d(\obsVec_i,\code) &= \min _{\codeVec \in \code} d_{\text{H}}(\obsVec_i,\codeVec).
\end{align}
Then, the MDCD problem can be formulated as follows.
\vspace{0.1cm}
\begin{problem}{\textsc{Minimum Distance Code Detection} (MDCD)}
	\textbf{Input:} Positive integers $N, n$, a binary $N \times n$ matrix $\obsMat$, and a set $\codeSet$ of binary linear codes of dimension $k\leq n$, where each $\code \in \codeSet$ is given by a generator matrix $\genMat$.

	\textbf{Output:} A generator matrix $\genMat$ of a binary linear code $\code_{\text{MDCD}} \in \codeSet$ such that:
	\begin{align}
		\code_{\text{MDCD}} = & \arg \min _{\code \in \codeSet} \sum _{i=1}^Nd(\obsVec_i,\code), \label{eq:mdcr}
	\end{align}
	where potential ties are broken arbitrarily.
\end{problem}
\vspace{0.1cm}

\textsc{Maximum Likelihood Code Detection} (MLCD) is a closely related problem that is of particular interest because it minimizes the detection error rate when all codes in $\codeSet$ are equiprobable. Let us assume that transmission takes place over a BSC with crossover probability $p \in \left(0,\frac{1}{2}\right)$, which we denote by BSC$(p)$, and let $d_{\text{H}}(\mathbf{a},\mathbf{b})$ denote the Hamming distance between $\mathbf{a}$ and $\mathbf{b}$. The MLCD problem, which was derived in~\cite{Valembois2001}, can then be formulated as follows.

\vspace{0.1cm}
\begin{problem}{\textsc{Maximum Likelihood Code Detection} (MLCD)}
	\textbf{Input:} Positive integers $N, n$, a binary $N \times n$ matrix $\obsMat$, and a set $\codeSet$ of binary linear codes of dimension $k\leq n$, where each $\code \in \codeSet$ is given by a generator matrix $\genMat$.

	\textbf{Output:} A generator matrix $\genMat$ of a binary linear code $\code_{\text{MLCD}} \in \codeSet$ such that:
	\begin{align}
		\code_{\text{MLCD}} = & \arg \max _{\code \in \codeSet} \prod _{i=1}^N\sum _{\codeVec \in \code} \left(\frac{p}{1-p}\right)^{d_{\text{H}}(\obsVec_i,\codeVec)}, \label{eq:mlcd}
	\end{align}
	where potential ties are broken arbitrarily.
\end{problem}
\vspace{0.1cm}

We discuss the relation between the MDCD problem and the MLCD problem in more detail in Section~\ref{sec:open}.

\section{The MDCD Problem for $|\codeSet| = \ell$}\label{sec:mdcd}
In this section, we prove that when we are given a fixed set of $\ell$ binary linear codes, finding a code that minimizes the sum distance from the noisy codewords is NP-hard. By a fixed set, here we mean a set of size which is constant in the input parameters, which is a restriction that can be added to the input of the formal definition of the MDCD problem. Typically, when studying the computational complexity of a problem, we refer to \emph{decision problems}, i.e., problems for which the answer is either ``yes'' or ``no''. In contrast, the MDCD problem defined above is an \emph{optimization problem}, i.e., a problem in which we are looking for a solution that optimizes an objective function, potentially under some constraints. However, the definition of NP-hardness can be extended to optimization problems using \emph{Turing reductions}, e.g., see the discussion on the complexity of search problems in~\cite[Chapter 5]{Garey1979}. We avoid talking about NP-completeness here intentionally, because the notion is only well-defined for the decision versions of the problems.

\begin{theorem}\label{thm:NPhard}
	The MDCD problem for $|\codeSet|=\ell$ is NP-hard.
\end{theorem}

We construct a reduction from the \textsc{Minimum Distance Decoding} problem (MDD), proven to be NP-hard in \cite{Berlekamp1978}.\footnote{The MDD problem was referred to as the \textsc{Coset Weights} problem in \cite{Berlekamp1978}, where it was defined as a decision problem. We reduce from the optimization version of the MDD problem which is NP-hard as well, since the objective function is computable in polynomial time~\cite[Chapter 5]{Garey1979}.} The MDD problem can be formulated as follows.
\vspace{0.2cm}
\begin{problem}{\textsc{Minimum Distance Decoding} (MDD)}
	\textbf{Input:} A generator matrix $\genMat$ of a binary linear code $\code$ of length $n$ and an $n$-bit binary vector $\mathbf{y}$.

	\textbf{Output:} An $n$-bit binary vector $\hat{\codeVec} = \arg\min_{\codeVec \in \code} d_H(\mathbf{y},\codeVec)$.
\end{problem}
\vspace{0.2cm}

Our reduction constructs an algorithm $\mathcal{A}_{\text{MDD}}$ that solves the MDD problem when given access to any algorithm $\mathcal{A}_{\text{MDCD}}$ that solves the MDCD problem. The algorithm $\mathcal{A}_{\text{MDD}}$ only makes a polynomial number of calls to $\mathcal{A}_{\text{MDCD}}$ and only performs polynomial-time computations otherwise. Therefore, if an efficient algorithm for MDCD existed, $\mathcal{A}_{\text{MDD}}$ would solve the MDD problem in polynomial time, which is not possible (unless $\text{P}=\text{NP}$) since the MDD problem is NP-hard.

\begin{algorithm}[t]
	\KwData{Full-rank $k \times n$ generator matrix $\genMat$, an $n$-bit binary vector $\mathbf{y}$.}
	\KwResult{Codeword $\hat{\codeVec} = \arg\min_{\codeVec \in \code} d_H(\mathbf{y},\codeVec)$.}
	$\genMat^{(k)}=\genMat$\;
	$l=k$\;
	\While{$l >0$}{
		$\{\genMat_1,\genMat_2,\genMat_3\} = \textsc{SplitCover}(\genMat^{(l)})$\;\label{algline:MDCD4}
		$\genMat^{(l-1)} = \mathcal{A}_{\text{MDCD}}(\mathbf{y},\{\genMat_1,\genMat_2,\genMat_3\})$\; \label{algline:MDCD5}
		$l = l-1$\;
		}
	$\hat{\codeVec} = \genMat^{(0)}$\; \smallskip
	\caption{Algorithm $\mathcal{A}_{\text{MDD}}$ for solving the MDD problem using $\mathcal{A}_{\text{MDCD}}$ as a subroutine.}\label{alg:2}
\end{algorithm}

More precisely, the $\mathcal{A}_{\text{MDCD}}$ algorithm has inputs $\codeSet$ (i.e., a set of $\ell$ generator matrices, here we take $\ell=3$) and the observation matrix $\obsMat$, and it outputs a generator matrix $\genMat$ for a code $\code \in \codeSet$ which is a solution to the MDCD problem. Our algorithm for solving the MDD problem using $\mathcal{A}_{\text{MDCD}}$ is given in Algorithm~\ref{alg:2}. The main idea is that, starting from the code $\genMat$ of dimension $k$ given as an input to the MDD problem, we call the \textsc{SplitCover} function described in Algorithm \ref{alg:3}. This function constructs (in polynomial time) three generator matrices $\genMat_1$, $\genMat_2$, and $\genMat_3$ of binary linear codes of dimension $(k-1)$, with the property that a codeword is generated by $\genMat$ if and only if it is generated by at least one of $\genMat_1$, $\genMat_2$, or $\genMat_3$. Then, we use the $\mathcal{A}_{\text{MDCD}}$ algorithm on $\mathbf{y}$ (i.e., the input of the MDD problem) and $\{\genMat_1,\genMat_2,\genMat_3\}$, which returns the code of dimension $(k-1)$ with the minimum distance from $\mathbf{y}$ that contains the solution to the MDD problem. We repeat this another $(k-1)$ times until the resulting code contains a single codeword, which is the solution $\hat{\codeVec}$ to the MDD problem.

In the following lemma, we prove the aforementioned properties of the \textsc{SplitCover} function.
\begin{lemma}\label{lemma:splitcover}
	\textsc{SplitCover} given in Algorithm~\ref{alg:3} takes an $l \times n$ matrix $\genMat$ of rank $l$ as an input and produces (in polynomial time) a set of three $(l-1)\times n$ generator matrices $\{\genMat_1,\genMat_2,\genMat_3\}$ with the following properties:
	\begin{enumerate}
		\item The rank of $\genMat_1$, $\genMat_2$, and $\genMat_3$ is $(l-1)$.
		\item $\text{span}(\genMat) = \text{span}(\genMat_1) \cup \text{span}(\genMat_2) \cup \text{span}(\genMat_3)$.
	\end{enumerate}
\end{lemma}
\begin{IEEEproof}
	The construction of $\genMat_1$, $\genMat_2$, and $\genMat_3$ is a concatenation of a subset of rows of $\genMat$, so it clearly has polynomial complexity. Moreover, by assumption, $\genMat$ has $l$ linearly independent rows. Since $\genMat_1$ and $\genMat_2$ are constructed using $(l-1)$ distinct rows of $\genMat$, they are clearly of rank $(l-1)$. Similarly, $\genMat_3$ is constructed using $(l-2)$ distinct rows of $\genMat$ and one row that is the sum of the remaining $2$ rows of $\genMat$, so it also clearly of rank $(l-1)$ and the first property follows. Finally, recall that $\text{span}(\genMat) = \left\{\mathbf{u}\genMat: \mathbf{u} \in \{0,1\}^k\right\}$. Since $\genMat_1$ is $\genMat$ with the second row omitted, it is easy to see that
	\begin{align}
		\text{span}(\genMat_1) & = \left\{\mathbf{u}\genMat: \mathbf{u} \in \{0,1\}^{k}, u_2 = 0\right\}.
	\end{align}
	Similarly, we have:
	\begin{align}
		\text{span}(\genMat_2) & = \left\{\mathbf{u}\genMat: \mathbf{u} \in \{0,1\}^{k}, u_1 = 0\right\}.
	\end{align}
	Finally, since the first row of $\genMat_3$ is equal to $(\genVec_1 + \genVec_2)$, $\text{span}(\genMat_3)$ will contain all vectors $\mathbf{u}\genMat$ for which either $u_1=0$ and $u_2=0$, or $u_1 = 1$ and $u_2 = 1$, or equivalently:
	\begin{align}
		\text{span}(\genMat_3) = \left\{\mathbf{u}\genMat: \mathbf{u} \in \{0,1\}^{k}, u_1 = u_2\right\}.
	\end{align}
	Since the set $\text{span}(\genMat_1) \cup \text{span}(\genMat_2) \cup \text{span}(\genMat_3)$ covers all possibilities for $u_1$ and $u_2$ and the remaining elements of $\mathbf{u}$ are free variables in all three cases, the second property follows.
\end{IEEEproof}

\begin{algorithm}[t]
	\SetKwProg{Fn}{Function}{:}{}
	\KwData{Full-rank $l \times n$ matrix $\genMat = \begin{bmatrix} \genVec_1^T  & \genVec_2^T  & \hdots &  \genVec_l^T	\end{bmatrix}^T$.}
	\KwResult{Set of three $(l-1) \times n$ matrices $\{\genMat_1,\genMat_2,\genMat_3\}$.}
	\Fn{\textsc{SplitCover}{$(\genMat)$}}{
		$\genMat_{1}	 = \begin{bmatrix}
		\genVec_1^T  & \genVec_3^T  & \hdots &  \genVec_l^T
		\end{bmatrix}^T$\;
		$\genMat_{2}	 = \begin{bmatrix}
		\genVec_2^T  & \genVec_3^T  & \hdots  & \genVec_l^T
		\end{bmatrix}^T$\;
		$\genMat_{3}	 = \begin{bmatrix}
		 (\genVec_1 + \genVec_2)^T & \genVec_3^T  & \hdots &  \genVec_l^T
		\end{bmatrix}^T$\; \smallskip
	\Return{$\{\genMat_1,\genMat_2,\genMat_3\}$}\;
	}\medskip
	\caption{Algorithm {\textsc{SplitCover}.}}\label{alg:3}
\end{algorithm}

\begin{IEEEproof}[Proof of Theorem \ref{thm:NPhard}]
	First, note that in our reduction, the observation matrix $\obsMat$ is in fact an $n$-bit binary vector and, in particular, it is the $n$-bit binary vector $\mathbf{\mathbf{y}}$ that is given as input to the MDD problem. In that case, the solution to the MDCD problem is a code $\code \in \codeSet$ such that:
	\begin{align}
			\code & = \arg\min_{\code \in \codeSet} d(\mathbf{y},\code) = \arg\min_{\code \in \codeSet} \left(\min_{\codeVec \in \code} d_H(\mathbf{y},\codeVec)\right),
	\end{align}
	where the last equation follows from the definition of $d(\mathbf{y},\code)$.

	Let $\mathcal{G}_{\ell} = \{\genMat_1,\hdots,\genMat_{\ell}\}$ denote a set of $\ell$ generator matrices and let $\text{span}(\mathcal{G}_{\ell}) = \bigcup _{i=1}^{\ell} \text{span}(\genMat_i)$. Then, identifying a code in $\mathcal{G}_{\ell}$ that is closest to $\mathbf{y}$ in terms of the minimum Hamming distance is equivalent to identifying a code in $\mathcal{G}_{\ell}$ that contains a codeword $\hat{\codeVec}= \arg\min_{\codeVec \in \text{span}(\mathcal{G}_{\ell})} d_{H}(\mathbf{y},\codeVec)$. In Algorithm~\ref{alg:2}, at every iteration $l$ it holds that $\text{span}(\genMat^{(l)}) = \text{span}(\genMat_1) \cup \text{span}(\genMat_2) \cup \text{span}(\genMat_3)$ by Lemma~\ref{lemma:splitcover}. By the discussion above and since we started from $\genMat^{(k)} = \genMat$, at every iteration $l$ of Algorithm~\ref{alg:2}, the $\mathcal{A}_{\text{MDCD}}$ algorithm identifies the code $\genMat^{(l-1)} \in \{\genMat_1, \genMat_2, \genMat_3\}$ that contains a solution $\hat{\codeVec}$ to the MDD problem. Since $\genMat^{(0)}$ is a single $n$-bit binary vector, Algorithm~\ref{alg:2} terminates by returning $\hat{\codeVec}$.

	Both $\mathcal{A}_{\text{MDCD}}$ and \textsc{SplitCover} are called $k$ times in Algorithm~\ref{alg:2}. Moreover, by Lemma~\ref{lemma:splitcover} we know that the complexity of \textsc{SplitCover} is polynomial. Finally, all remaining computations can clearly be carried out in polynomial time, meaning that the overall complexity of our reduction is polynomial.
\end{IEEEproof}
One can view our reduction as a ternary search-style procedure, where the space of all codewords is split into three sets (which only have a constant overlap of codewords) and the set containing a solution is returned by the $\mathcal{A}_\text{MDCD}$ algorithm.

\section{The MDCD Problem for $\codeSet=\mathcal{LC}_k$}\label{sec:valembois}

In Section~\ref{sec:mdcd}, we studied the MDCD problem when $\codeSet$ is a fixed set of $\ell$ binary linear codes. In contrast, in~\cite{Valembois2001} the author formulated the MDCD problem when $\codeSet$ is the space of all possible linear codes of a given dimension $k$, which we will denote by $\mathcal{LC}_k$. We note that the MDCD problems for $\codeSet = \mathcal{LC}_k$ and for $\codeSet = \{\code_1,\code_2,\hdots,\code_\ell\}$ are fundamentally different. When $\codeSet=\mathcal{LC}_k$, we are looking for \emph{some} code among all possible linear codes that minimizes the total distance from the noisy codewords and there might be a very large number of codes that are solutions to the problem. On the other hand, when $\codeSet = \{\code_1,\code_2,\hdots,\code_\ell\}$, we need to decide which code is closest to the observation matrix $\obsMat$ in terms of the minimum Hamming distance, which might be a much harder task to do.

In \cite{Valembois2001}, it is stated that the MDCD problem is equivalent\footnote{Such an equivalence result would indeed imply that the MDCD problem is generally NP-hard when $\codeSet=\mathcal{LC}_k$, which is the claim attributed to~\cite{Valembois2001} in certain related works~(e.g., \cite{Chabot2007,Carrier2018}). However, the equivalence statement appears without proof in~\cite{Valembois2001}.} to a \textsc{Rank Reduction} (RR) problem, which is then proven to be NP-hard via a reduction from the \textsc{Minimum Distance} problem~\cite{Vardy1997}. The term ``rank-reduction'' already hints at the fact that such an equivalence requires that the rank of the observation matrix $\obsMat$ is at least $k$, which implies that at least $k$ noisy codewords have to be observed. However, in the practical application described in Section~\ref{sec:intro}, the number of observations (and thus  $\textrm{rank}(\obsMat)$) is always significantly smaller than $k$, since the decision latency and the signal processing cost have to be minimized. 

In this case, it turns out that it is simple to identify the computational complexity of the MDCD problem. In particular, we describe a polynomial-time algorithm that can find a code $\code \in \mathcal{LC}_k$ that minimizes $\sum _{i=1}^{N}d(\obsVec_i,\code)$ when $\textrm{rank}(\obsMat) \leq k$. The main idea of the algorithm is that, since $\textrm{rank}(\obsMat) \leq k$, we can always construct a full-rank $k \times n$ generator matrix $\genMat$ with $\obsMat$ as a submatrix to achieve $\sum_{i=1}^N d(\obsVec_i,\code)=0$ in polynomial time. This algorithm has two steps: the first step ensures that $\sum_{i=1}^N d(\obsVec_i,\code)$ is minimized, while the second step ensures that $\genMat$ has rank $k$ and thus generates a code of the desired dimension $k$.

\begin{algorithm}[t]
 \KwData{Full-rank $r \times n$ generator matrix $\genMat$ from step 1.}
 \KwResult{Full-rank $k \times n$ generator matrix $\genMat$.}
 $i = 1$\;
 \While{$\textrm{rank}(\genMat) < k$ and $i \leq n$}{
	$\genMat' = \begin{bmatrix} \genMat \\ \mathbf{e}_i \end{bmatrix}$\;
  \If{$\textrm{rank}(\genMat') > \textrm{rank}(\genMat) $}{
   $\genMat = \genMat'$\;
   }
	 $i = i + 1$\;
 }
 \caption{Rank augmentation of $\genMat$.}\label{alg:1}
\end{algorithm}%

\textbf{Step 1:} Let $\textrm{rank}(\obsMat) = r \leq k$ and let $\mathcal{L} = \left\{l_1, l_2, \hdots, l_r \right\}$ denote a set of indices of any $r$ linearly independent rows of $\obsMat$. The set $\mathcal{L}$ can be constructed in polynomial time using Gaussian elimination. We construct the $r$ first rows of $\genMat$ as:
\begin{align}
	\genMat_{r \times n}	& = \begin{bmatrix}
		\obsVec_{l_1}^T & \hdots & \obsVec_{l_r}^T
	\end{bmatrix}^T.
\end{align}

\textbf{Step 2:} Let $\mathbf{e}_i$ denote the standard basis row vector of length $n$ with a $1$ in the $i$-th coordinate and $0$'s elsewhere. We extend $\genMat$ to have dimensions $k \times n$ and rank $k$ by following the procedure of Algorithm \ref{alg:1}. This procedure is guaranteed to construct a full-rank $k \times n$ generator matrix $\genMat$ and it requires at most $n$ steps, with each step having polynomial complexity. The final $k \times n$ generator matrix $\genMat$ has the following form:
\begin{align}
	\genMat_{k \times n}	& = \begin{bmatrix}
			\obsVec_{l_1}^T & \hdots & \obsVec_{l_r}^T & \mathbf{e}_{i_1}^T & \hdots & \mathbf{e}_{i_{k-r}}^T
	\end{bmatrix}^T,
\end{align}
for some $\{i_{1},\hdots,{i_{k-r}}\} \subset \{1,\hdots,n\}$. Since the $2^k$ codewords of the code $\code$ corresponding to $\genMat$ are generated as $\mathbf{u}\genMat$, where $\mathbf{u} \in \{0,1\}^k$, it is easy to see that $\obsVec_i \in \code, \, \forall i = 1,\hdots,N$. This means that $\sum_{i=1}^N d(\obsVec_i,\code) = 0$ and $\code$ indeed minimizes $\sum_{i=1}^N d(\obsVec_i,\code)$.


\section{Open Problems}\label{sec:open}
In this section, we identify and discuss some interesting open problems related to the complexity of code detection.

\subsection{Computational Complexity of the MLCD Problem}
Unlike minimum distance decoding and maximum likelihood decoding which are equivalent over the BSC (and known to be NP-hard~\cite{Berlekamp1978}), MDCD is generally not equivalent to MLCD. This is demonstrated through the following example.
\begin{example}
Consider the case where transmission takes place over a BSC$(0.25)$, we have $|\codeSet| = 2$, and the full-rank generator matrices $\genMat_1$ and $\genMat_2$ that describe the codes $\code_1$ and $\code_2$ (both of dimension $k=3$), respectively, are:
\begin{align}
	\genMat_1 = \begin{bmatrix}
		 0 & 1 & 0 & 0 & 1 \\
     1 & 1 & 1 & 0 & 0 \\
     1 & 1 & 1 & 1 & 1
	\end{bmatrix},
	\;
	\genMat_2 = \begin{bmatrix}
		 0 & 1 & 0 & 1 & 0 \\
     1 & 0 & 0 & 1 & 0 \\
     0 & 1 & 1 & 0 & 0
	\end{bmatrix}.
\end{align}
Moreover, let us assume that we have the following observation matrix with a single noisy codeword $\obsMat = \obsVec_1 = \begin{bmatrix} 1 & 1 & 1 & 0 & 0 \end{bmatrix}$. Finally, let us define:
\begin{align}
	f(\code) & = \sum _{\codeVec \in \code} \left(\frac{p}{1-p}\right)^{d_{\text{H}}(\obsVec_1,\codeVec)},
\end{align}
so that $\code_{\text{MLCD}} = \arg\max_{C \in \{\code_1,\code_2\}} f(\code)$.
It is easy to verify that $d(\obsVec_1,\code_1) = 0$ and $d(\obsVec_1,\code_2) = 2$, but $f(\code_1) = 1.449$ and $f(\code_2) = 1.481$, meaning that $\code_{\text{MDCD}} \neq \code_{\text{MLCD}}$. So, the code that is the optimal solution of the MDCD problem is not the optimal solution of the MLCD problem, and vice-versa.
\end{example}

In~\cite{Valembois2001}, it is not explained rigorously how MDCD is related to MLCD. Here, we provide the following explanation. Let $\alpha = \frac{p}{1-p}$. Then, using the well-known max-log approximation with a base-$\alpha$ logarithm and the fact that $\log _{\alpha}(x)$ is decreasing in $x$ since $\alpha \leq 1$, we can re-write \eqref{eq:mlcd} as:
\begin{align}
	\code_{\text{MLCD}} = & \arg \max _{\code \in \codeSet} \sum _{i=1}^N \log _{\alpha}\left(\sum _{\codeVec \in \code} \alpha^{d_{\text{H}}(\obsVec_i,\codeVec)}\right) \\
					\approx & \arg \min _{\code \in \codeSet} \sum _{i=1}^N \min _{\codeVec \in \code}d_{\text{H}}
(\obsVec_i,\codeVec) =\code_{\text{MDCD}}.
\end{align}
Note however, that this approximation does not imply anything about the computational complexity of MLCD merely from the computational complexity of MDCD, nor vice-versa.

Arguably, the maximum likelihood objective of the MLCD problem is a better distance metric than the minimum distance objective of the MDCD problem, since it minimizes the probability of detection error. As such, studying the complexity of the MLCD problem is an important next step. In this direction, one could attempt to construct a reduction from the MDD problem to the MLCD problem by replacing $\mathcal{A}_{\text{MDCD}}$ with an algorithm $\mathcal{A}_{\text{MLCD}}$ that solves the MLCD problem in Algorithm~\ref{alg:2}. However, for this to work one would have to show that the code $\genMat ^{(l-1)}$ returned by $\mathcal{A}_{\text{MLCD}}$ always contains the solution to the MDD problem (as shown for $\mathcal{A}_{\text{MDCD}}$ in the proof of Theorem~\ref{thm:NPhard}), which does not necessarily hold.

\subsection{Detection Complexity for Subclasses of Linear Codes}
Similarly to the case of maximum likelihood decoding, it would be interesting to examine specific subclasses of linear codes (e.g., LDPC codes, polar codes), for which, in principle, efficient algorithms for the MDCD problem could exist. In this direction, given a subclass of linear codes, our reduction can be applied if this subclass is \emph{closed} under a \emph{split-cover} operation similar to \textsc{SplitCover} defined in Algorithm~\ref{alg:3}. Specifically, closure in this context means that a full-rank $k \times n$ generator matrix $\genMat$ that belongs to the given subclass of linear codes, can be split into $\ell$ full-rank $(k-1)\times n$ generator matrices $\genMat_1,\hdots,\genMat_\ell$, that belong to the same subclass such that $\text{span}(\genMat) = \bigcup _{i=1}^{\ell} \text{span}(\genMat_i)$. A procedure that generates $\genMat_1,\hdots,\genMat_\ell$, in polynomial time can be then used instead of the specific \textsc{SplitCover} function that we used in Algorithm~\ref{alg:2} in order to prove hardness for specific subclasses of codes.

\subsection{Complexity of MDCD for any $\ell$ and $N$}
The proof of Theorem~\ref{thm:NPhard} establishes the NP-hardness of the MDCD problem when $\ell=3$ and $N=1$, which is sufficient to show that the problem is NP-hard in general. 

A very similar reduction can be used to prove NP-hardness for any $\ell > 3$. The main idea is that Lemma~\ref{lemma:splitcover} can be extended to the case where $\genMat$ is split into $\ell$ distinct\footnote{Note that, if the codes in $\codeSet$ are not required to be distinct, the NP-hardness of the MDCD problem with $\ell > 3$ follows easily from the NP-hardness of the $\ell = 3$ case since we can simply set, e.g., $\genMat_\ell = \genMat_3$ for all $\ell > 3$.} codes $\genMat_1,\hdots,\genMat_\ell$. We note that the case of $\ell = 1$ is trivial and the NP-hardness of the case when $\ell = 2$ follows easily from the NP-hardness of the case when $\ell = 3$. Specifically, a hypothetical polynomial-time algorithm for the $\ell = 3$ case could call a hypothetical polynomial-time algorithm for the $\ell = 2$ case three times (one for each of the three possible pairs of candidate codes) and combine the partial results in order to solve the MDCD problem.

The case where $N>1$ observations are available is also of practical interest. Showing NP-hardness for a given $N>1$ is an open problem, which does not seem to follow directly from the techniques we have used in this work.

\section{Conclusion}
In this work, we studied the fundamental problem of the computational complexity of code detection for binary linear codes and we proved that the MDCD problem is NP-hard through a reduction from the \textsc{Minimum Distance Decoding} problem in the practically relevant case where $\codeSet$ contains a fixed number $\ell$ of candidate codes. Moreover, we identified a number of open problems, the most
interesting being the computational complexity of the MLCD problem.

\section{Acknowledgment}
The work of Alexios Balatsoukas-Stimming is supported by the Swiss National Science Foundation project \#175813. The work of Aris Filos-Ratsikas is supported by the Swiss National Science Foundation under contract No. 200021\_165522. The authors would like to thank the anonymous reviewers for their useful suggestions.

\bibliographystyle{IEEEtran}
\bibliography{IEEEabrv,ISIT2019}

\end{document}